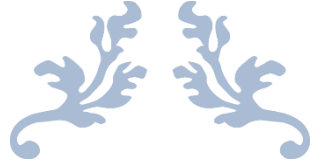

# Analyzing the impact of two major factors on medical expenses paid by health insurance organization in Iran

1. **Seyed Nasser Moosavi (corresponding author)**

(Master's degree of industrial engineering, Yazd University, Yazd, Iran)

Cell NO: +989138216471

Email: Seyednasserm6@gmail.com, s.n.mosavi@stu.yazd.ac.ir

2. **Ashkan Khalifeh**

(Professor, Department of Statistics, Yazd University, Yazd, Iran)

3. **Ali Shojaee**

(Doctor, Member of the board of directors of Iranian Health Insurance Organization (IHIO), Tehran, Iran – MD, MPH)

4. **Masoud Abessi**

(Professor of Information Systems, Department of industrial engineering, Yazd university, Yazd, Iran)

**Abstract**

In healthcare scope, the profound role of insurance companies is undeniable. Health insurance establishment's main responsibility is to support public health financially and promote the quality of health services. Government's subsidies to healthcare insurance, insured payments and insurance companies' costs must be specified in such a way that both people and insurers mutually benefit. In this research, we propose a model for determining healthcare costs paid by health insurance organization with regard to two major factors, the geographical regions where the patients live and the seasons when they receive service, using two-way ANOVA method. Since both effects are found to be significant (P-value=0.00<0.01), allocating different insurance costs to the people residing in different regions, and also changing the patterns of insurance extension in different seasons with regard to the results derived from the research, can detract healthcare costs and give more satisfaction to the lower income insured patients.

Keywords: health insurance, healthcare cost, geographical region, season, two-way ANOVA

**1. Introduction**

Nowadays, the visions associated with healthcare are diverse and service quality expectations are on the rise. The public growing concerns about the safety, quality of healthcare services, and payment equality in healthcare expenses demand a responsive healthcare system. Although in recent years, Iran's healthcare system has achieved outstanding improvements in the public health level and indices pertaining to that, one of the main concerns of the government is how to create an equilibrium between receiving and paying for healthcare service all over the country. One alternative would be to revise the policies and tariffs helping provide public health insurances.

Jones et al.[1] studied the changes in insurance coverage of women of childbearing age after implementing Care Act. They concluded that coverage has improved for many groups of society. However, some gaps in coverage remain for Latinas and some other states. Lee[2] determined the factors affecting the total medical expense for depression patients in the emergency room, using a multiple regression. These factors consisted of gender, age, main illness, course of admission to the emergency room and length of stay. Mpofu et al.[3] studied associations between non-communicable disease (NCD) risk factors, race, education and insurance using chi-square tests and multivariate logistic regression. Hsiao and Cheng[4] analyzed the association between socio-economic status of the patients and the hospital care they receive, under a universal health insurance scheme in Taiwan. The results proved that lower education results in poorer perceived quality of care. Lemos et al.[5] studied the costs relating to multiple and singleton births in the US. They concluded that pregnancies with the delivery of twins and triplets or more cost almost 5 and 20 times as much respectively, when compared with singleton pregnancies. Ibrahimipour et al.[6] investigated three main tasks of health financing system (revenue collection, risk pooling and purchasing) using interviews in Iran. They showed seven major obstacles to universal coverage. Kataoka et al.[7] studied the association of unmet need (defined as having a need for mental health evaluation but not using any service in a 1-year period) with ethnicity and insurance status, using three national household surveys. Singh et al.[8] reviewed health insurance coverage for the people in Uttar Pradesh (UP), India. The results revealed that health insurance schemes cover only 4.8% population, which is not satisfactory. Son et al.[9] carried out a research involving the data from 2008 to 2011 living profiles of older people survey and representing the relationship between frailty and medical expenses in community-dwelling elderly patients. They proved that frailty is a predictor of increasing medical expenses. Chen et al.[10] tried to examine whether the quantity of medication received by the patients with diabetes and hypertension in Taiwan is associated with desired healthcare outcomes and expenses. They came to prove that undersupply or oversupply of medications resulted in poorer healthcare outcomes, and also oversupply results in higher total healthcare expenses. Cheng[11] evaluated advantages and disadvantages associated with a single-payer, national health insurance scheme administered by the government's Department of Health and recently replacing a patchwork of separate social health insurance funds in Taiwan. Hurd et al.[12] investigated how health insurance influences the use of healthcare services by the elderly. They indicated a positive association between the health insurance and healthcare service utilization. Cummins et al.[13] compared two marketing channels through which the property-liability insurance is distributed. They found that the independent agency system is less efficient than the exclusive agency system. Using a natural experiment in Taiwan, Chiao et al.[14] determined a significant association between the availability of National Health Insurance (NHI) and longitudinal increase in life satisfaction among older adults. An[15] studied the effects of obesity and smoking on consumption of substantial healthcare resources by U.S. adults aged 18 years and older from 1998 to 2011. They indicated a positive association.

Healthcare in Iran is based on three pillars: the public government system, the private sector, and NGOs. In April 2014, the first phase of a new health plan (Tarh-e-salaamat) covering up to 90 percent of costs for patients' medical bills at public hospitals was introduced. Since this plan tries to integrate all the public and private insurers, the main challenge it faces is to make insurers support the process during which the plan is implemented. Some experts believe that if this plan is not supported enough and the integration principle is not performed in a correct way, it will result in the fatal consequences. Having considered all the resources and limitations, it was approved to detract the people's out-of-pocket costs to 30 percent of the whole remedial expense. On the other hand, insurance companies claim that the money paid by the people is rational and fair, and detracting their share is not economical for the insurers. In this paper, we study the patients' expenses paid by the insurance organization during 5 consecutive years, and analyze the effect of geographical location where the patients live and the season when they visit the healthcare centers on the expense paid by the insurance organization. Comparing the costs paid by the people in different seasons and regions of Iran can give us a good wisdom on how to allocate health expenses to the people and insurers in different regions and seasons and also helps the government determine whether the shares specified by the plan are rational or not.

## 2. Methods and materials

In this paper, the data relating to 827637 patients who have been in the hospitals under contract to insurance organization to receive medical services have been analyzed. We are going to scrutinize the effects of two factors, say, the geographical regions where the patients live and the seasons during which they receive healthcare service on the service expense that the insurers undertake to cover. The response variable is the healthcare cost paid by the insurance organization and is denoted by "Payment" in analysis. The values of this variable are given in Iran currency. In order to analyze the effects of two mentioned factors on the response variable, a two-way ANOVA technique is applied. Since the number of patients receiving healthcare service varies in different seasons and geographical regions, an unbalanced ANOVA design is used. Also, the two factors in the model are fixed-effect factors, because all of their levels are considered in analysis. We have used IBM SPSS (version 22) software to process the data.

## 3. Results

433381 of the patients studied in this paper are male and 394256 of them are female. The factor season has four levels, and the frequency of the patients receiving service in different levels of this factor is shown in Table 1.

Table 1. The frequency of the patients in different levels of factor season

| SEASON | FREQUENCY | PAYMENT MEAN | PAYMENT SD |
|---|---|---|---|
| SPRING | 210313 | 3463044.82 | 6120371.247 |
| SUMMER | 234198 | 3874440.31 | 6702892.094 |
| AUTUMN | 209809 | 4002731.49 | 6955923.718 |
| WINTER | 173317 | 3966727.92 | 6906003.517 |
| TOTAL | 827637 | 3821764.48 | 6673198.530 |

The next factor to be considered is the geographical region the patients reside in. As indicated in Figure 1, Iran has 31 provinces which we have divided into 4 major regions, due to the different climates and geographical locations. Region 1 (Northern Iran) is comprised of 7 provinces such as Tehran, the capital of Iran and the provinces adjacent to Caspian Sea. This region has a mild and rather damp climate. Region 2 (Central Iran) consists of 7 provinces and has a mild and dry climate. Region 3 (Western Iran) is comprised of 11 western provinces. These provinces are located in a mountainous region where Zagros mountains are stretched from north to south. And region 4 (Eastern Iran) is formed of 6 provinces located in central and eastern parts of the country. The major part of this area is covered with deserts and the climate is dry and hot. Two known deserts of this area are Dasht-e Loot and Dasht-e Kavir.

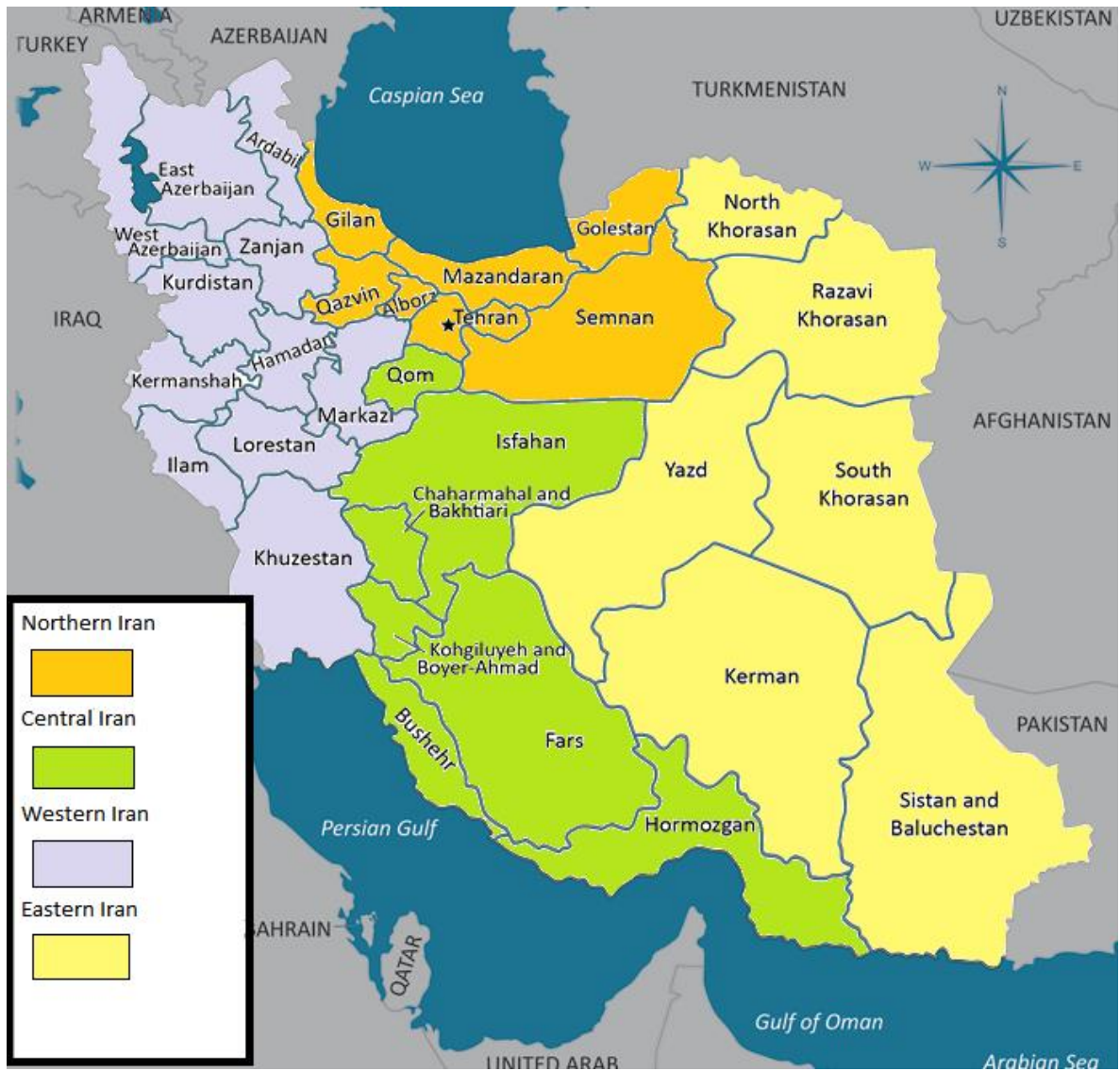

Fig. 1. The major geographical regions where the patients live

Table 2 represents the frequency of the patients in these regions. Since the number of the patients significantly varies among the Region's levels, a proportionate sampling has been applied here.

Table 2. The frequency of the patients in different levels of factor Region

| REGION | FREQUENCY | PAYMENT MEAN | PAYMENT SD |
|---|---|---|---|
| NORTHERN IRAN | 659011 | 3479354.29 | 6229837.213 |
| CENTRAL IRAN | 29996 | 5235595.64 | 7880399.533 |
| WESTERN IRAN | 123835 | 5204830.76 | 8118541.116 |
| EASTERN IRAN | 14795 | 4630874.70 | 7671503.583 |
| TOTAL | 827637 | 3821764.48 | 6673198.530 |

Two basic assumptions must hold to validate the analysis. The first one is normality of the residuals. Since F statistic is used to examine whether the whole model and factors' effects are significant, the results derived from the model will not be reasonable if the normality assumption is violated. Table 3 indicates one-sample Kolmogorov-Smirnov test examining normality of the residuals. Considering α=0.01, the null hypothesis stating the residuals have normal distribution can't be rejected.

Table 3. Testing normality of residuals, using one-sample Kolmogorov-Smirnov test

|  |  | RESIDUAL FOR PAYMENT |
|---|---|---|
| N |  | 827629 |
| NORMAL PARAMETERS | mean | 0 |
|  | Std. deviation | 6635000.68790 |
| TEST STATISTIC |  | 0.008 |
| ASYMP. SIG. |  | 0.020 |

The second basic assumption to be scrutinized is equality of the residuals' variances. Figure 2 represents different values of residuals against predicted values of the response variable, levels of factor season and levels of factor region. The variation of residuals for each variable is almost changeless and we don't need to be worried of wrong conclusions. Also, the residuals don't show any specific and clear pattern.

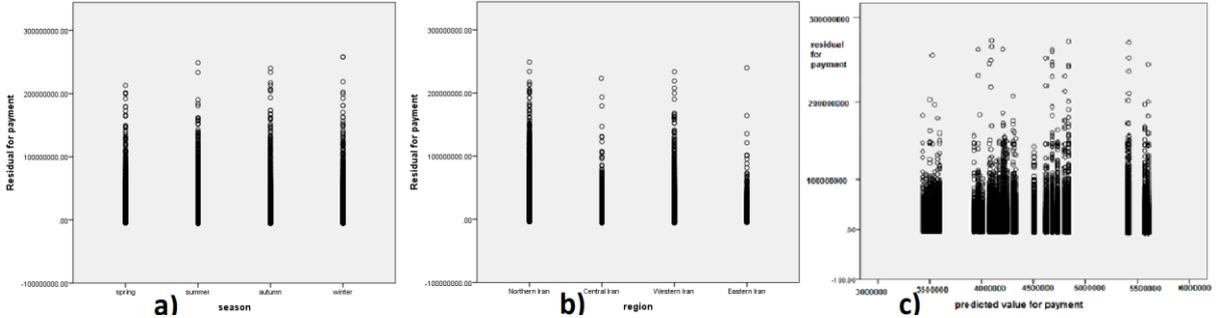

Fig. 2. Plotting residuals versus a) season, b) region, and c) predicted response variable, to test the variance equality of residuals

Finally, Table 4 indicates the significance of the whole model and also separate and simultaneous effects of the two factors on dependent variable, payment.

Table 4. Testing the significance of the whole model and the factors' effects

| SOURCE | SUM OF SQUARES | DEGREES OF FREEDOM | MEAN SQUARE | F | SIG. | NONCENT. PARAMETER | OBSERVED POWER |
|---|---|---|---|---|---|---|---|
| CORRECTED MODEL | 4.207E+17 | 15 | 2.805E+16 | 637.063 | 0 | 9555.946 | 1 |
| INTERCEPT | 3.072E+17 | 1 | 3.072E+18 | 69777.7 | 0 | 69777.7 | 1 |
| REGION | 3.723E+17 | 3 | 1.241E+17 | 2819.075 | 0 | 8457.226 | 1 |
| SEASON | 5.052E+15 | 3 | 1.684E+15 | 38.251 | 0 | 114.752 | 1 |
| REGION*SEASON | 2.36E+15 | 9 | 2.622E+14 | 5.956 | 0 | 53.608 | 1 |
| ERROR | 3.644E+19 | 827621 | 4.402E+13 | | | | |
| TOTAL | 4.894E+19 | 827637 | | | | | |
| CORRECTED TOTAL | 3.686E+19 | 827636 | | | | | |

The results represented in this table are as follows:

i- The whole model is significant (P-Value=0.000<0.01). It means that at least one of the separate or simultaneous effects is significant.

ii- The separate effect of region on payment is significant (P-Value=0.000<0.01). It means that the healthcare cost in different regions is not the same, in other words, the null hypothesis $H_0$: $\mu_1=\mu_2=\mu_3=\mu_4$ in which $\mu_i$ is the mean of healthcare cost paid by the insurer in $i^{th}$ region is rejected, and at least two means don't equal each other. On the rest of the paper, we will study these differences more precisely.

iii- The separate effect of season on healthcare cost is significant (P-Value=0.000<0.01). Thus, the mean of healthcare cost is not the same in different seasons.

iv- The interaction of two factors is also significant (P-Value=0.000<0.01). It means that the impact of one factor hinges upon the level of the latter.

Since healthcare costs of the patients in different seasons and regions are not the same, we are interested in scrutinizing these differences. Using Scheffe's Post-Hoc test, Table 5 compares each level of factor Region with its other levels and also categorizes different regions based on the mean of healthcare costs paid for the patients residing in different regions.

Table 5. Comparing healthcare costs in different geographical regions

| REGION (I) | REGION (J) | MEAN DIFFERENCE (I-J) | STD. ERROR | SIG. | %95 CONFIDENCE INTERVAL | | HOMOGENEOUS SUBSETS |
|---|---|---|---|---|---|---|---|
| | | | | | LOWER BOUND | UPPER BOUND | |
| Northern Iran | Central Iran | -1756241.36 | 39172.341 | .000 | -1865747.21 | -1646735.50 | 1 |
| | Western Iran | -1725476.48 | 20550.190 | .000 | -1782924.31 | -1668028.64 | |
| | Eastern Iran | -1151520.42 | 55158.104 | .000 | -1305714.30 | -997326.53 | |
| Central Iran | Northern Iran | 1756241.36 | 39172.341 | .000 | 1646735.50 | 1865747.21 | 3 |
| | Western Iran | 30764.88 | 42698.662 | .915 | -88598.77 | 150128.53 | |
| | Eastern Iran | 604720.94 | 66657.947 | .000 | 418379.37 | 791062.51 | |
| Western Iran | Northern Iran | 1725476.48 | 20550.190 | .000 | 1668028.64 | 1782924.31 | 3 |
| | Central Iran | -30764.88 | 42698.662 | .915 | -150128.53 | 88598.77 | |
| | Eastern Iran | 573956.06 | 57715.855 | .000 | 412612.01 | 735300.11 | |
| Eastern Iran | Northern Iran | 1151520.42 | 55158.104 | .000 | 997326.53 | 1305714.30 | 2 |
| | Central Iran | -604720.94 | 66657.947 | .000 | -791062.51 | -418379.37 | |
| | Western Iran | -573956.06 | 57715.855 | .000 | -735300.11 | -412612.01 | |

Table 6 compares healthcare costs with regard to the season when the patients receive the service, and indicates homogeneous subsets for factor season.

Table 6. Comparing healthcare costs in different seasons

| SEASON (I) | SEASON (J) | MEAN DIFFERENCE (I-J) | STD. ERROR | SIG. | %95 CONFIDENCE INTERVAL | | HOMOGENEOUS SUBSETS |
|---|---|---|---|---|---|---|---|
| | | | | | LOWER BOUND | UPPER BOUND | |
| Spring | summer | -411300.58 | 19932.525 | .000 | -467021.74 | -355579.42 | |
| | autumn | -539677.50 | 20473.321 | .000 | -596910.45 | -482444.55 | 1 |
| | winter | -503603.52 | 21525.257 | .000 | -563777.15 | -443429.90 | |
| Summer | spring | 411300.58 | 19932.525 | .000 | 355579.42 | 467021.74 | |
| | autumn | -128376.92 | 19945.134 | .000 | -184133.33 | -72620.51 | 2 |
| | winter | -92302.94 | 21023.515 | .000 | -151073.95 | -33531.93 | |
| Autumn | spring | 539677.50 | 20473.321 | .000 | 482444.55 | 596910.45 | |
| | summer | 128376.92 | 19945.134 | .000 | 72620.51 | 184133.33 | 3 |
| | winter | 36073.98 | 21536.934 | .423 | -24132.29 | 96280.24 | |
| Winter | spring | 503603.52 | 21525.257 | .000 | 443429.90 | 563777.15 | |
| | summer | 92302.94 | 21023.515 | .000 | 33531.93 | 151073.95 | 3 |
| | autumn | -36073.98 | 21536.934 | .423 | -96280.24 | 24132.29 | |

## 4. Discussion

As shown in Table 5, the separate effects and interactions of factors are statistically significant. Considering factor region, it is apparent that the mean of healthcare cost in northern provinces is profoundly less than in other provinces of the country (P-Value=0.000<0.01). Eastern provinces have the least cost after northern provinces (P-Value=0.000<0.01). The costs of central and western provinces do not differ significantly (P-Value=0.915>0.05) and they have the highest means of healthcare cost amongst all. The last column of the table indicates homogeneous subsets for different regions. Three major subgroups have been defined for factor Region. As indicated in Table 6 representing healthcare costs with regard to the season when the patients receive service, healthcare costs in autumn and winter aren't significantly different, and also the highest costs are for these two seasons. The healthcare costs of different seasons can be categorized as shown in the last column of the table.

The factors considered in this research have proved to profoundly affect healthcare costs. To get a better perception of these effects, Figure 3 represents the Mean of healthcare costs in different regions and seasons. The least healthcare cost relates to northern region and spring and the highest cost is for central region and summer. As mentioned before, since the lines in figure 3 cross each other in several points, the interaction is significant.

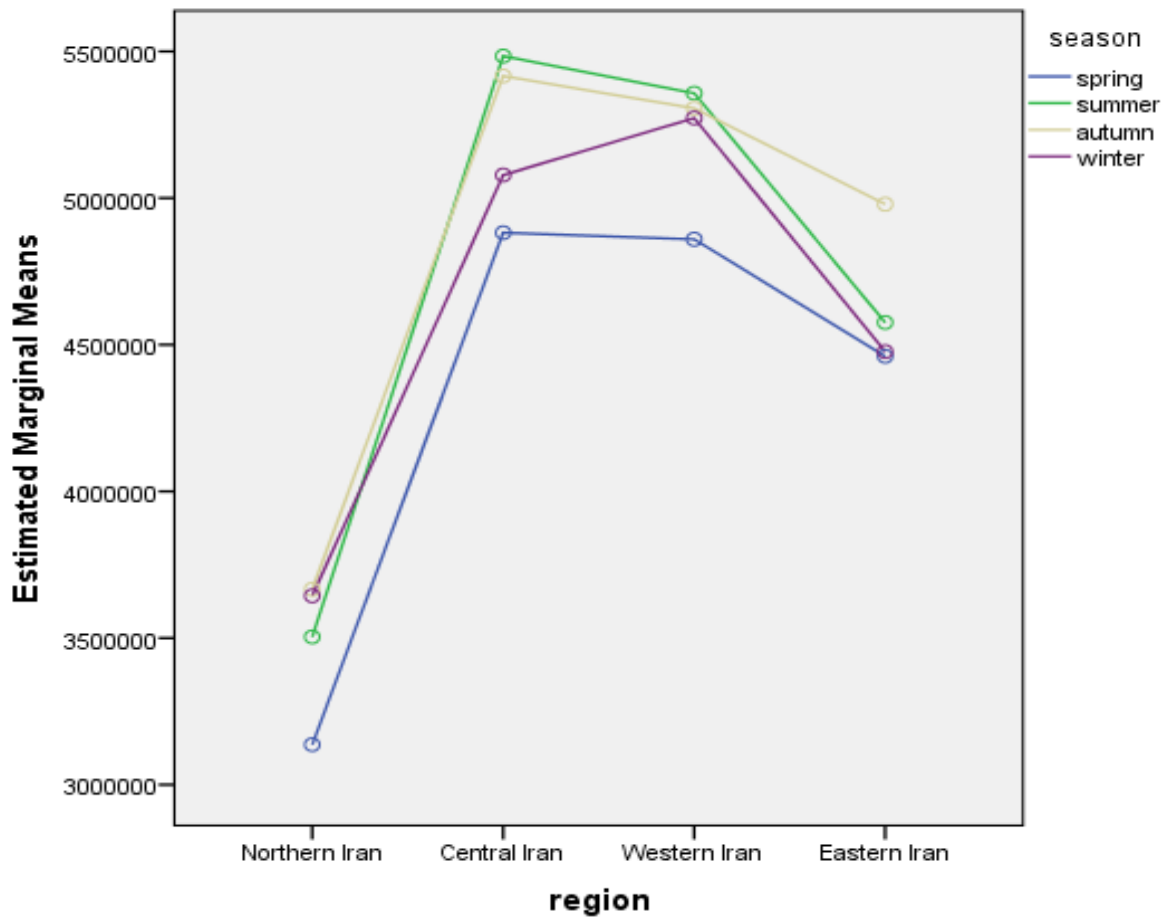

Fig 3. Estimated marginal means of payment for different levels of two associated factors

Many factors including social, financial and medical ones affecting healthcare costs have been scrutinized so far, but none of the previous studies has analyzed the effects of two factors considered here, so this work offers a new finding.

All in all, in this research, we studied how the season when the patients receive medical service and the geographical regions where they live affect the healthcare costs paid by the insurers. Since these two factors are recognized to be effective, allocating different insurance costs to the people residing in different regions, and also changing the patterns of insurance extension in different seasons can detract healthcare costs and give more satisfaction to the lower income patients.

Since the recognition of these two factors is really easy and doesn't need any specific measuring tool, it's useful and economical to make use of this model for predicting healthcare costs and offering insurance extension patterns. However, there are other social, economic and medical factors whose effects on healthcare costs are yet to be well studied and deserve extensive research. A drawback of this model is that we have assumed that no patient visits the hospitals out of their own city, while this assumption doesn't hold in reality. Moreover, some patients are hospitalized for a period of more than three months and we consider only their entry date.